%% file: main.tex
\documentclass[pra,aps,twocolumn,groupedaddress,floatfix,preprintnumbers,nofootinbib]{revtex4-1}
\usepackage{amsmath,amssymb,multirow,epsfig,bm,color}
\usepackage{graphicx}

\newcommand{\beq}{\begin{equation}}
\newcommand{\eeq}{\end{equation}}
\newcommand{\bea}{\begin{eqnarray}}
\newcommand{\eea}{\end{eqnarray}}

\newcommand{\eF}{\varepsilon_{F}}
\newcommand{\pF}{p_{F}}
\newcommand{\kF}{k_{\textrm{F}}}
\newcommand{\vF}{v_{\textrm{F}}}

\newcommand{\grad}{\boldsymbol{\nabla}} 
\newcommand{\meff}{M_{\textrm{eff}}}
\newcommand{\abs}[1]{\lvert{#1}\rvert}            

\begin{document}

\title{
Spin-polarized vortices with reversed circulation
}

\author{Piotr Magierski$^{1,2}$}\email{piotr.magierski@pw.edu.pl}
\author{Gabriel Wlaz\l{}owski$^{1,2}$}\email{gabriel.wlazlowski@pw.edu.pl}
\author{Andrzej Makowski$^{1}$}
\author{Konrad Kobuszewski$^{1}$}

\affiliation{$^1$Faculty of Physics, Warsaw University of Technology, Ulica Koszykowa 75, 00-662 Warsaw, Poland}
\affiliation{$^2$Department of Physics, University of Washington, Seattle, Washington 98195--1560, USA}

\begin{abstract} 
We present the analysis of the structure of fermionic vortices with the spin-polarized core from a weak coupling limit to the unitary regime.
We show the mechanism for the generation of the {\it reversed circulation} in the vortex core induced by an excess of majority spin particles.
We introduce the classification of the polarized vortices based on the number of Fermi circles where the minigap vanishes.
This provides a unique description of the vortex as one cannot smoothly map wave functions into one another corresponding to vortices differing by the number of Fermi circles.
The effective mass of quasiparticles along the vortex core is analyzed and its role in the propagation of spin-polarization along the vortex line is discussed.
\end{abstract}


\maketitle
\section{Introduction} 
Quantum vortices are one of the most prominent examples of topological excitations in superfluids~\cite{Simula-Book,vortices2002}.
They occur both in bosonic systems, where ${}^4$He liquid below lambda point and atomic BECs are prime examples, as well as in fermionic systems including superfluid ${}^3$He, metallic superconductors or fermionic ultracold gases.
They are also believed to exist in superfluid neutron matter forming neutron stars.
Although the stability of the vortex originates from the topology of the order parameter, its properties vary significantly for fermionic and bosonic systems.
Namely, in bosonic systems at low temperatures, the core of the vortex is essentially empty as the superfluid density reaches zero in the center of the vortex.
The only particles that can reside there are those which form the thermal cloud vanishing at T=0~\cite{zng}.
In the case of fermionic systems, the strength of the interparticle interaction to large extent defines the core structures (see e.g.~Refs.~\cite{salomaa1987, gygi1991, nygaard2003, prem2017} discussing the vortex structures in ${}^3$He, II-type superconductors, fermionic ultracold gases and in multiply quantized vortices, respectively).

For dilute Fermi gases the interaction is parametrized via dimensionless quantity $a\kF$, where $a$ is $s$-wave scattering length and $\kF=(3\pi^2n)^{1/3}$ is Fermi wave vector corresponding to the density $n$.
If $a\kF$ is positive then bound states (dimers) are formed, and typical characteristics of bosonic systems are recovered, with the modification that bosons can split into two fermions, which may form a normal state occupying the center of the vortex.
In the far BEC limit ($a\kF\rightarrow 0^+$) this would require significant excitation energy and therefore in practice is not expected to occur below the condensation temperature.
The situation is different for the dimers that are getting weakly bound when approaching the unitary limit ($a\kF\rightarrow \pm \infty$) at which their binding energy eventually reaches zero.
At a certain point, the first Andreev state appears inside the core and the density of normal fermions becomes nonzero in the core.
As the strength of the interaction becomes weaker the system enters into the BCS regime ($a\kF\rightarrow 0^-$) where fermions with opposite spins form Cooper pairs.
In this regime, the density of Andreev states increases, implying that density of matter in the core reaches a significant level, comparable with the bulk value~\cite{Machida2005,randeria2006,Machida2006}.

Spin imbalance may serve as another degree of freedom affecting pairing properties in Fermi system.
It also affects the structure of the vortex as the excess of unpaired fermions tend to accumulate at the core~\cite{takahasi2006,drummond2007}.
In this paper, we investigate impact of the spin polarization on the structure of the vortex in weakly and strongly interacting Fermi superfluid.
\begin{figure}[b]
   \begin{center}
   \includegraphics[width=0.99\columnwidth]{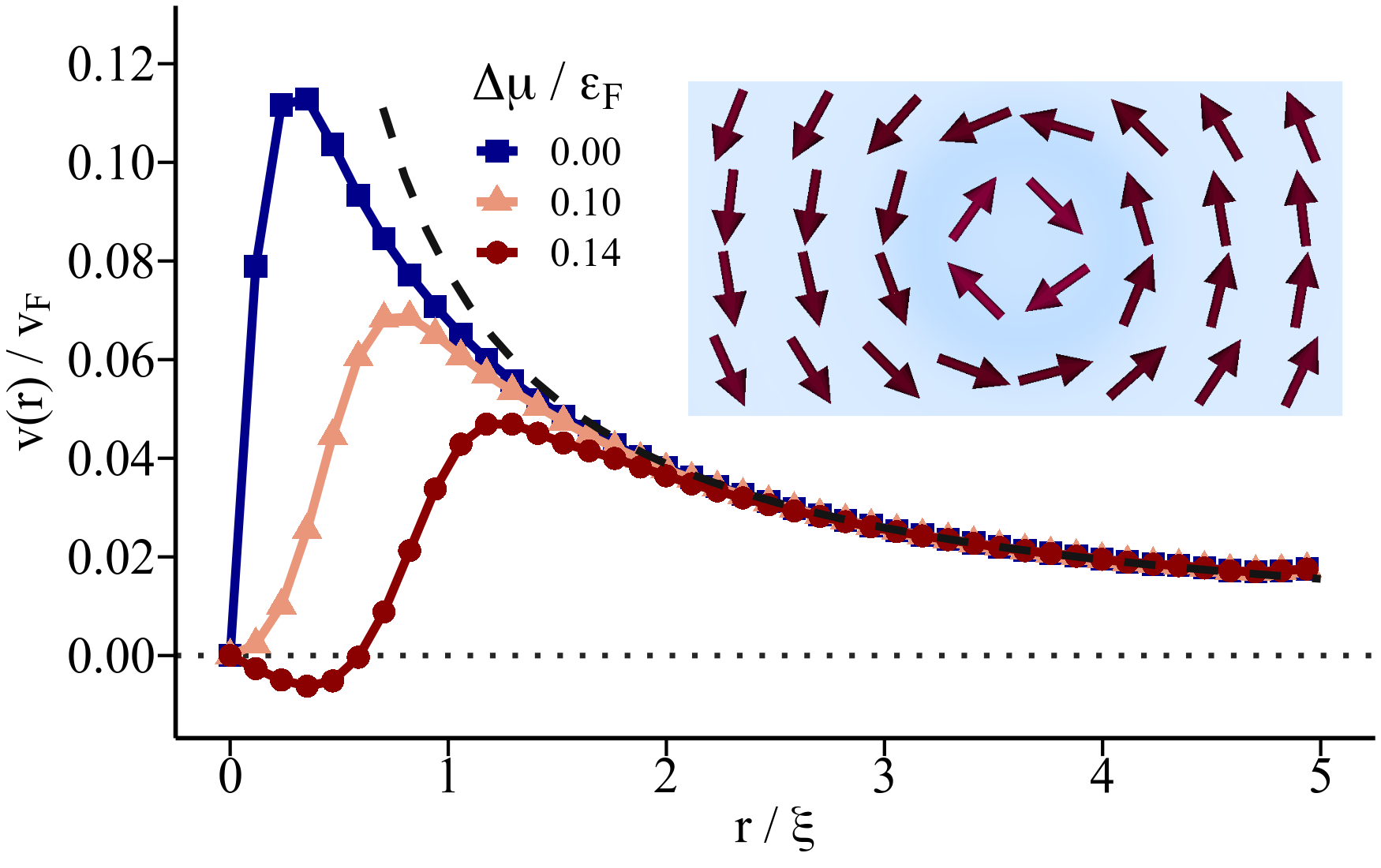} 
   \end{center}\vspace{-3mm}
   \caption{
    Evolution of velocity profiles (in units of the Fermi velocity $\vF=\pF/m$) as a function of distance from the core (in units of BCS coherence length $\xi$).
    Calculations were carried out for $a\kF=-0.84$ and 
    correspond to selected chemical potential differences of two spin states $\Delta\mu = \mu_{\uparrow}-\mu_{\downarrow}$.
    For sufficiently large spin imbalance the flow in the vortex core exhibits reversed circulation (red line, circles).
    Structure of the vortex core for such case is visualized in the inset: arrows show the direction of the flow, while the color map displays the density distribution of the fluid.
    Dashed line corresponds to velocity profile of an ideal quantum vortex $v(r)\sim 1/r$.
    }
   \label{fig:polarized-vortex}
\end{figure}
We find that the spin-imbalance affects the flow inside the vortex core, leading eventually to its inversion at sufficiently high imbalances.
This peculiar phenomenon is presented in Fig.~\ref{fig:polarized-vortex}, where we show velocity fields as a function of distance from the core.
The three cases correspond to different amounts of mismatch between chemical potentials of two spin components $\Delta\mu = \mu_{\uparrow}-\mu_{\downarrow}$.
As we increase $\Delta\mu$, the velocity field in the core is suppressed, and eventually changes direction.
In this letter we reveal the origin of the reversed circulation and discuss its consequences.
The effect is relevant for ultracold atomic systems with spin imbalance at the BCS regime up to the unitary limit, where quantum vortices were already observed~\cite{ZwierleinVortex,ZwierleinVortexImbalanced} and numerically simulated \cite{Wlazlowski2018}, and also to neutron stars.
Particularly for magnetars that are expected to generate magnetic field of the order or larger than $10^{16}$G~\cite{Turollaetal,BlaschkeChamel} which is sufficient to effectively spin polarize neutron matter inside vortex core~\cite{Steinetal2016, Pecak2021}.

\section{BdG equations for spin-imbalanced system}\label{sec:suppeq}
Our studies rely on Bogoliubov-de Gennes (BdG) formalism.
The explicit form of 
BdG equations for spin-imbalanced system reads (no spin-orbit coupling is considered):
\begin{align}\label{eq:hfbspin}
\begin{gathered}
{\cal H} 
\begin{pmatrix}
u_{n,\uparrow}(\bm{r}) \\
u_{n,\downarrow}(\bm{r}) \\
v_{n,\uparrow}(\bm{r}) \\
v_{n,\downarrow}(\bm{r})
\end{pmatrix}
= E_n
\begin{pmatrix}
u_{n,\uparrow}(\bm{r}) \\
u_{n,\downarrow}(\bm{r}) \\
v_{n,\uparrow}(\bm{r}) \\
v_{n,\downarrow}(\bm{r})
\end{pmatrix} \\
{\cal H} =  
\begin{pmatrix}
h_{\uparrow}(\bm{r}) -\mu_{\uparrow}  & 0 & 0 & \Delta(\bm{r}) \\
0 & h_{\downarrow}(\bm{r}) - \mu_{\downarrow}& -\Delta(\bm{r}) & 0 \\
0 & -\Delta^*(\bm{r}) &  -h^*_{\uparrow}(\bm{r}) +\mu_{\uparrow} & 0 \\
\Delta^*(\bm{r}) & 0 & 0& -h^*_{\downarrow}(\bm{r}) + \mu_{\downarrow}
\end{pmatrix}
\end{gathered}
\end{align}
where $\mu_{\uparrow,\downarrow}$ are chemical potentials for two spin components.
Single particle hamiltonian in the BdG approximation is defined as $h_{\uparrow}=h_{\downarrow}=-\frac{\hbar^2}{2m}\nabla^2$.
The form of the Hamiltonian leading to the BdG equations reads:
\begin{eqnarray}
&\hat{H}&=
\sum_{\sigma=\uparrow,\downarrow}\int d\bm{r}\;
\hat{\psi}_{\sigma}^\dagger(\bm{r})
\left[-\frac{\hbar^2}{2m}\nabla^2-\mu_\sigma\right]
\hat{\psi}_\sigma(\bm{r}) \nonumber \\
& &+\frac{g}{2} \sum_{\sigma=\uparrow,\downarrow}
\int d\bm{r}
\hat{\psi}_{ \sigma}^\dagger\left(\bm{r}\right)
\hat{\psi}_{-\sigma}^\dagger\left(\bm{r}\right)
\hat{\psi}_{-\sigma}        \left(\bm{r}\right)
\hat{\psi}_{ \sigma}        \left(\bm{r}\right)
\end{eqnarray}
with coupling constant $g$.
In the BdG equations one usually omit the mean-field term contributing to $h_{\uparrow}$ and $h_{\downarrow}$ and  takes into account pairing contribution $\Delta_{\uparrow\downarrow}(\bm{r})=\Delta(\bm{r})=g\langle\hat{\psi}_{\downarrow} \left(\bm{r}\right)\hat{\psi}_{ \uparrow}\left(\bm{r}\right) \rangle$ only. Then the formalism is applicable to weakly interacting (BCS) regime.
In more general case, the single particle hamiltonian $h_{\sigma}$
explicitly depends on the spin state.
For example, asymmetric superfluid local density approximation (ASLDA), that applies to the unitary Fermi gas (UFG), provides $h_{\sigma} = \nabla\frac{-\hbar^2}{2m^*_{\sigma}(p)}\nabla + U_{\sigma}(n,p)$, where $m^*_{\sigma}$ is an effective mass of particle with spin $\sigma=\{\uparrow,\downarrow\}$ that depends on local polarization $p(\bm{r})=\frac{n_{\uparrow}(\bm{r})-n_{\downarrow}(\bm{r})}{n_{\uparrow}(\bm{r})+n_{\downarrow}(\bm{r})}$, and $U$ is a mean field which depends on the polarization and the total density of particles $n(\bm{r})=n_{\uparrow}(\bm{r})+n_{\downarrow}(\bm{r})$.
For explicit form of the ASLDA energy density functional and corresponding single particle hamiltonian see Ref.~\cite{LNP__2012}.
The pairing gap is related to quasi-particle wave-functions:
\begin{equation}
\Delta(\bm{r}) = -\dfrac{g_{\textrm{eff}}}{2}\sum_{0<E_n<E_c} (u_{n,\uparrow}(\bm{r})v_{n,\downarrow}^{*}(\bm{r})-u_{n,\downarrow}(\bm{r})v_{n,\uparrow}^{*}(\bm{r})),\\
\end{equation}
where $g_{\textrm{eff}}$ is a regularized coupling constant and $E_c$ is cut-off energy scale, see~\cite{LNP__2012} for details of the regularization scheme.
In the mean-field BdG approximation, the coupling constant is related to the scattering length (bare coupling constant is given by $g=4\pi\hbar^2 a/m$) whereas for ASLDA the coupling constant is fitted to the quantum Monte Carlo data.
The densities $n_{\sigma}$ and currents $\bm{j}_{\sigma}$ of spin components are constructed as:
\begin{eqnarray}
n_{\sigma}(\bm{r}) &=& \sum_{0<E_n<E_c}\abs{v_{n,\sigma}(\bm{r})}^2\\
\bm{j}_{\sigma}(\bm{r}) &=& \sum_{0<E_n<E_c} \textrm{Im}[v_{n,\sigma}(\bm{r})\nabla v_{n,\sigma}^*(\bm{r})] 
\label{eqn:densities0}
\end{eqnarray}

The BdG equations~(\ref{eq:hfbspin}) decouple into two independent sets:
\begin{align}\label{eq:hfbspin1}
\begin{gathered}
\begin{pmatrix}
h_{\uparrow}(\bm{r})-\mu  &  \Delta(\bm{r}) \\
\Delta^*(\bm{r}) &  -h^*_{\downarrow}(\bm{r})+\mu
\end{pmatrix} 
\begin{pmatrix}
u_{n,\uparrow}(\bm{r}) \\
v_{n,\downarrow}(\bm{r})
\end{pmatrix}
= E_{n+}
\begin{pmatrix}
u_{n,\uparrow}(\bm{r}) \\
v_{n,\downarrow}(\bm{r})
\end{pmatrix} ,
\end{gathered}
\end{align}
\begin{align}\label{eq:hfbspin2}
\begin{gathered}
\begin{pmatrix}
h_{\downarrow}(\bm{r})-\mu& -\Delta(\bm{r}) \\
-\Delta^*(\bm{r}) &  -h^*_{\uparrow}(\bm{r})+\mu
\end{pmatrix} 
\begin{pmatrix}
u_{n,\downarrow}(\bm{r}) \\
v_{n,\uparrow}(\bm{r})
\end{pmatrix}
= E_{n-}
\begin{pmatrix}
u_{n,\downarrow}(\bm{r}) \\
v_{n,\uparrow}(\bm{r})
\end{pmatrix},
\end{gathered}
\end{align}
where $\mu=\frac{1}{2}(\mu_{\uparrow}+\mu_{\downarrow})$ denotes mean chemical potential and $E_{n\pm}=E_n \pm \frac{\Delta\mu}{2}$ with $\Delta\mu=\mu_{\uparrow}-\mu_{\downarrow}$.
Solutions of equations (\ref{eq:hfbspin1}) and (\ref{eq:hfbspin2}) are connected via symmetry relation, namely if vector $\varphi_{+}=\left( u_{n\uparrow},v_{n\downarrow}\right)^{T}$ represents a solution of Eq.~(\ref{eq:hfbspin1}) with eigenvalue $E_n$, then vector $\varphi_{-}=(v_{n\uparrow}^*,u_{n\downarrow}^*)^{T}$ is a solution of Eq.~(\ref{eq:hfbspin2}) with eigenvalue $-E_n$.
In practice it is sufficient to solve  equations (\ref{eq:hfbspin1}) only (for all quasiparticle energy states), and then solutions with positive quasiparticle energies contribute to the spin-down densities, whereas solutions with negative energies to the spin-up densities.

\begin{table*}[]
\hspace{-1.1cm}
\begin{tabular}{c|cccccc|ccc|}
\cline{2-10}
                                                         & \multicolumn{6}{c|}{BCS}                                                                                                                                                                 & \multicolumn{3}{c|}{UFG}                                                         \\ \hline
\multicolumn{1}{|c|}{$P=\left|\tfrac{N_\uparrow-N_\downarrow}{N_\uparrow+N_\downarrow}\right|$[\%]}                           & \multicolumn{1}{c|}{0.0}                  & \multicolumn{1}{c|}{0.5} & \multicolumn{1}{c|}{1.0} & \multicolumn{1}{c|}{0.0}                  & \multicolumn{1}{c|}{0.5} & 1.0 & \multicolumn{1}{c|}{0.0}                  & \multicolumn{1}{c|}{0.5} & 1.0 \\ \hline
\multicolumn{1}{|c|}{Lattice}                            & \multicolumn{3}{c|}{150x150x32}                                                                       & \multicolumn{3}{c|}{100x100x80}                                                  & \multicolumn{3}{c|}{100x100x120}                                                 \\ \hline
\multicolumn{1}{|c|}{$k_F$}                              & \multicolumn{3}{c|}{1.222}                                                                            & \multicolumn{3}{c|}{0.756}                                                       & \multicolumn{3}{c|}{0.510}                                                       \\ \hline
\multicolumn{1}{|c|}{$\Delta_\infty{[}\varepsilon_F{]}$} & \multicolumn{3}{c|}{0.06}                                                                             & \multicolumn{3}{c|}{0.16}                                                        & \multicolumn{3}{c|}{0.53}                                                        \\ \hline
\multicolumn{1}{|c|}{$\varepsilon_F$}                    & \multicolumn{3}{c|}{0.747}                                                                            & \multicolumn{3}{c|}{0.286}                                                       & \multicolumn{3}{c|}{0.130}                                                       \\ \hline
\multicolumn{1}{|c|}{$\xi {\,[}\Delta x{]}$}               & \multicolumn{3}{c|}{13.7}                                                                             & \multicolumn{3}{c|}{8.5}                                                         & \multicolumn{3}{c|}{3.7}                                                         \\ \hline
\multicolumn{1}{|c|}{$a k_F$}                            & \multicolumn{3}{c|}{-0.61}                                                                            & \multicolumn{3}{c|}{-0.84}                                                       & \multicolumn{3}{c|}{$\infty$}                                                    \\ \hline
\multicolumn{1}{|c|}{$\mu_\uparrow {[}\varepsilon_F{]}$}        & \multicolumn{1}{c|}{\multirow{2}{*}{1.031}} & \multicolumn{1}{c|}{1.077} & \multicolumn{1}{c|}{1.089} & \multicolumn{1}{c|}{\multirow{2}{*}{0.279}} & \multicolumn{1}{c|}{0.294} & 0.299 & \multicolumn{1}{c|}{\multirow{2}{*}{0.014}} & \multicolumn{1}{c|}{0.026} & 0.027 \\ \cline{1-1}
\multicolumn{1}{|c|}{$\mu_\downarrow {[}\varepsilon_F{]}$}        & \multicolumn{1}{c|}{}                       & \multicolumn{1}{c|}{0.986} & \multicolumn{1}{c|}{0.975} & \multicolumn{1}{c|}{}                       & \multicolumn{1}{c|}{0.265} & 0.260 & \multicolumn{1}{c|}{}                       & \multicolumn{1}{c|}{0.004} & 0.003 \\ \hline
\end{tabular}
\caption{Characteristic parameters used in the numerical calculations. $N_\sigma$ stands for particle number of given spin $\sigma$. The Fermi wave vector and Fermi energy are related to density at large distance from the vortex core $n_\infty$ as follows  $\kF=\sqrt{2m\eF}/\hbar=(3\pi^2n_{\infty})^{1/3}$. The $\Delta_\infty$ stands for the paring gap far from the vortex and defines the coherence length through relation $\xi=\eF/\kF\Delta_\infty$.  Chemical potentials for individual spin components are indicted as $\mu_\uparrow$ and $\mu_\downarrow$. }
\label{tab:setup}
\end{table*}
The equations were solved numerically for selected parameters presented in table~\ref{tab:setup}.
The calculations were executed on spatial 3D lattice  $N_x\times N_y \times N_z$ with lattice spacing $\Delta x$.
We considered straight vortex along $z$-direction, and thus by imposing the generic form of wave functions $\varphi(\bm{r})=\varphi(x,y)e^{ik_z z}$ the problem was effectively reduced to collections of 2D problems (parametrized by quantum number $k_z$).
For calculations in BCS regime we have applied BdG approximation, while for calculations at the unitarity ASLDA functional has been employed~\cite{LNP__2012}. The vortex solution was generated by imprinting technique, i.e. by imposing the particular structure of the order parameter 
of the form 
$\Delta(x,y)=|\Delta(\sqrt{x^2+y^2})|e^{i \phi}$ with $\phi=\arctan(y/x)$. The calculations have been performed using W-SLDA Toolkit~\cite{Wlazlowski2018,Bulgac2014,WSLDAToolkit}.

\begin{figure*}[]
   \begin{center}
  \includegraphics[width=.9\textwidth]{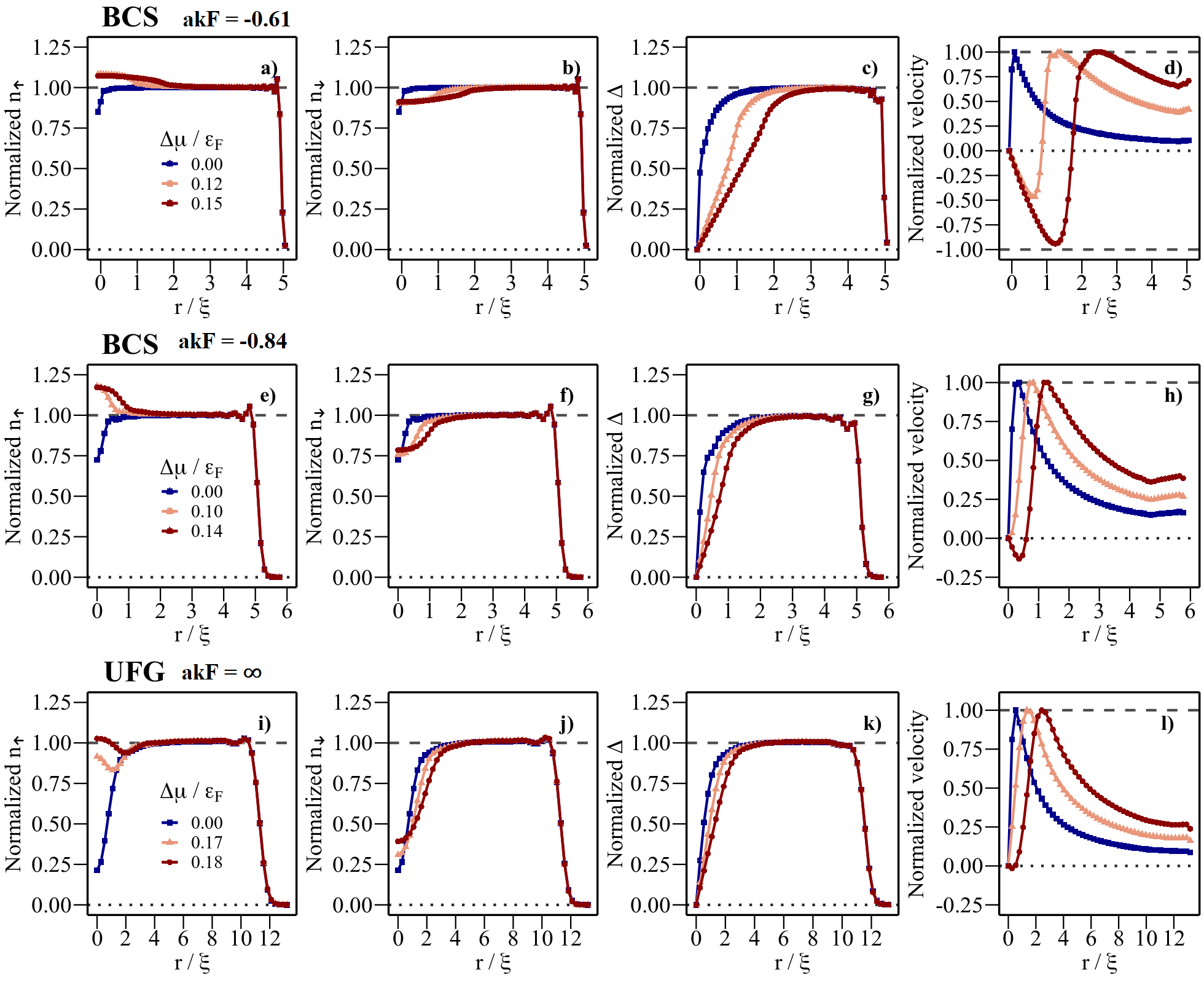}
   \end{center}\vspace{-3mm}
   \caption{
   Cross sections through the vortex core of various quantities presented as a function of radial direction for BCS (a-h) and UFG (j-l) regimes, respectively.
   Densities $n_{\uparrow,\downarrow}(r)$ (first two columns) and order parameter $|\Delta(r)|$ (third column) are normalized to their bulk values, while velocity field $v(r)$ is normalized to its maximal value (last column).
   The velocity profile is computed as $v(r)=(j_{\uparrow}(r)+j_{\downarrow}(r))/(n_{\uparrow}(r)+n_{\downarrow}(r))$.
   In each case, for the velocity profile we recover expected dependence $v(r)\sim 1/r$ for large distances (except regions where $n\rightarrow 0$, due to large numerical uncertainties). 
   }
   \label{fig:supplement}
\end{figure*}

Fig.~\ref{fig:supplement} presents cross sections through the vortex core along radial directions for 
the following quantities: spin-up  $n_{\uparrow}(r)$ and spin-down $n_{\downarrow}(r)$ atoms density, strength of the order parameter $|\Delta(r)|$ and the velocity field $v(r)=\abs{\bm{j}(\bm{r})}/n(\bm{r})$.
Different lines correspond to different spin-imbalance populations measured by chemical potential difference $\Delta\mu=\mu_{\uparrow}-\mu_{\downarrow}$.
The lattice formulation of the problem implies the
usage of periodic boundary conditions.
In order to remove the impact of periodicity on the results 
we placed the system in the potential well of a given 
radius $R$.
The external potential induces the vanishing of the 
density and the order parameter close to the boundary of simulation domain.
Clearly, the extracted velocity field is affected close to the boundary and thereby substantial numerical uncertainties occur in regions $n\rightarrow 0$. However here we focus on the core properties and the structure of the vortex in the
vicinity of the core is properly reproduced.
In partocular, the main effect that is the subject of 
the analysis is profoundly visible in Fig.~\ref{fig:supplement}(d): the reversed flow is present for spin-imbalanced solutions. 

\section{The origin of the reversed circulation} 
The properties of polarized vortices are determined by the states in the cores.
Their energies, for the unpolarized case, have been first estimated in Ref.~\cite{caroli1964}.
In the BCS limit, due to the separation of scales related to pairing (coherence length $\xi$) and single particle motion (de Broglie wavelength $\lambda_B$), these states can be conveniently described in the Andreev approximation~\cite{stone1996}.
In the unitary regime despite the fact that chemical potential $\mu$ is of the same order as the pairing gap $\Delta$, as will be seen below, it can still provide useful qualitative relations.

\begin{figure}[t]
   \begin{center}
   \includegraphics[width=0.99\columnwidth]{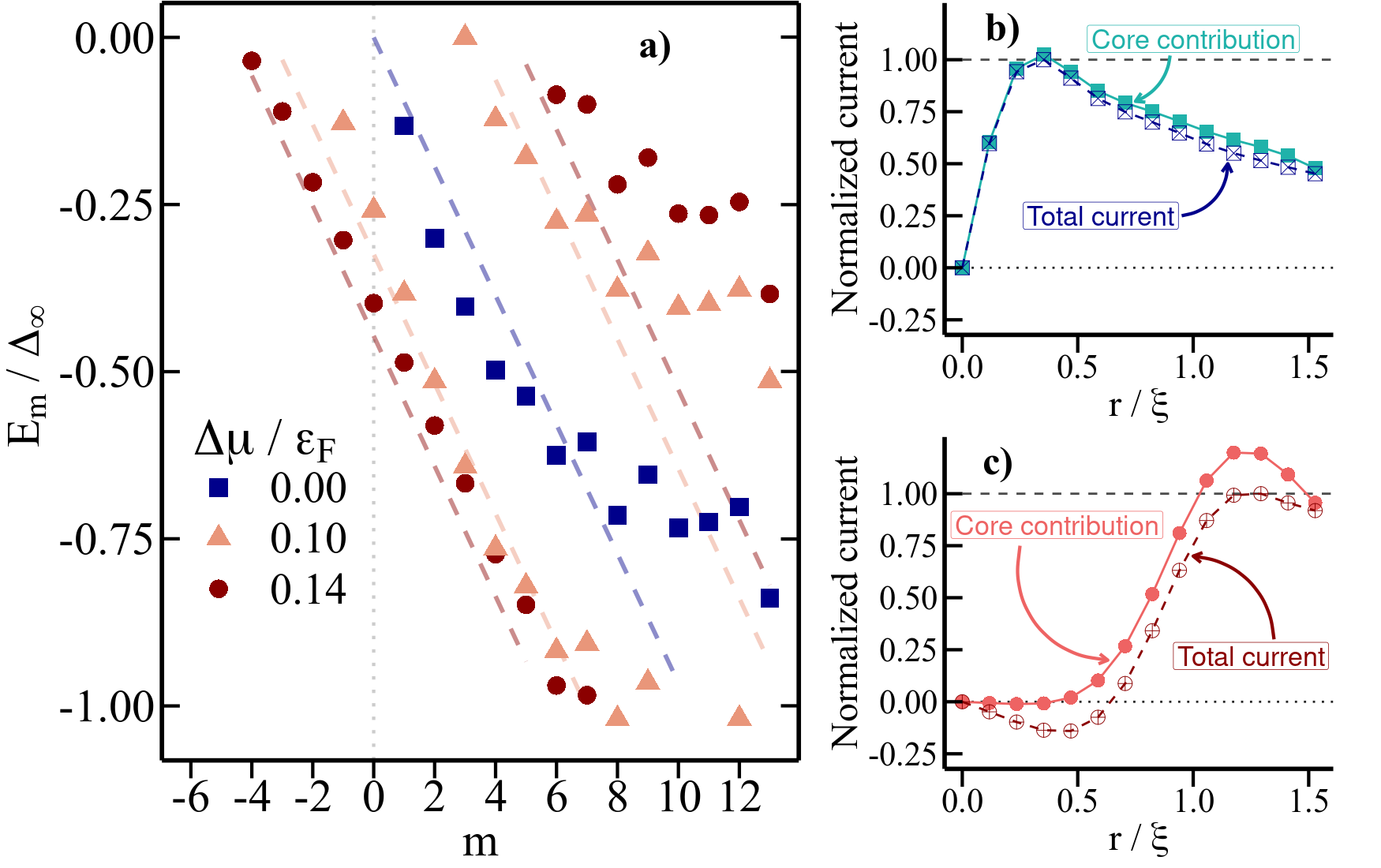} 
   \end{center}\vspace{-3mm}
   \caption{
   Quasiparticle energies $E_m$ of states in the core as a function of magnetic quantum number $m$ for selected spin imbalances of the system, panel a).
   The energies are expressed in units of paring field far away from the vortex core $\Delta_{\infty}$ (bulk value).
   The points are extracted as numerical solution of BdG equations for BCS regime $a\kF=-0.84$, while (dashed) lines display prediction of eq. (\ref{andreev_pol}) with $r_v=0.86 \xi$.
   The BdG equations were solved numerically in 3D, but only states with wave vector along the vortex line $k_z=0$ are shown (see section ~\ref{sec:suppeq} for details).
   Panels b) and c) display comparison of currents, generated by subgap states (filled symbols) and all states (open symbols): for spin-symmetric system $\Delta\mu/\eF=0$ (b) and for spin-imbalanced system  $\Delta\mu/\eF=0.14$ (c).
   In both case the currents are normalized to the maximal value of the corresponding total current.
   \label{fig:andreev-states}
   }
\end{figure}

In this approximation one decomposes the variation of $u$ and $v$ components of 
wave-functions (see Eq.~(\ref{eq:hfbspin})) at the Fermi surface into rapidly oscillating parts associated with $\kF$ and smooth variations governed by the coherence length, i.e.~$u(\bm{r})= e^{i\bm{k}_\textrm{F}\cdot\bm{r}}\tilde{u}(\bm{r})$ with $|\bm{k}_\textrm{F}|=\kF$, and similarly for the $v$ component~\cite{andreev1964}.
The Andreev approximation can be also used for studies of spin imbalance systems, providing the local polarization is relatively weak $\Delta\mu = \mu_{\uparrow}-\mu_{\downarrow} \ll \frac{1}{2}(\mu_{\uparrow}+\mu_{\downarrow}) \approx \eF= \kF^{2}/2 $.
For the reasons presented in Sec.~\ref{sec:suppeq} we will focus only on one set of BdG equations, which in Andreev
approximation describing states close to the Fermi surface acquire the form (we set $\hbar=m=1$):
\begin{equation} \label{bdg1}
 \left( \begin{array}{cc}
 -i\bm{k}_\textrm{F}\cdot\grad& \Delta(\textbf{r}) \\
 \Delta^*(\textbf{r}) & i\bm{k}_\textrm{F}\cdot\grad
\end{array} \right)\left( \begin{array}{cc}
\tilde{u}_{n,\uparrow}(\textbf{r}) \\
\tilde{v}_{n,\downarrow}(\textbf{r})
\end{array} \right) =\tilde{E}_{n+} \left( \begin{array}{cc}
\tilde{u}_{n,\uparrow}(\textbf{r}) \\
\tilde{v}_{n,\downarrow}(\textbf{r})
\end{array} \right),
\end{equation}
where $\tilde{E}_{n+} = E_{n+} + \frac{\Delta\mu}{2}$.
The second pair of equations for $\tilde{u}_{n,\downarrow}(\textbf{r})$ and $\tilde{v}_{n,\uparrow}(\textbf{r})$ has similar form and correspond to $\tilde{E}_{n-} = E_{n-} - \frac{\Delta\mu}{2}$. 
One may consider a schematic structure of a vortex core defined by the pairing field $\Delta(\bm{r})$ expressed in the polar variables ($\rho,\phi$): $\Delta(\rho,\phi)=|\Delta|e^{i\phi}\theta(\rho-r_{v})$ (counterclockwise rotating vortex), where $\theta$ is Heaviside step function.
Ignoring for the moment the degree of freedom along the vortex axis (2D case), one may solve eqs.~(\ref{bdg1}) and arrive at the quantization conditions associated with the trajectory of angular momentum $L_{z}$ (detailed derivation is provided in Appendix~\ref{appendixA}):
\begin{eqnarray}
\frac{\tilde{E}_{n+}}{\eF}\kF r_{v}\sqrt{1-\left (\frac{L_{z}}{\kF r_{v}} \right )^{2}}+\arccos{\left ( \frac{-L_{z}}{\kF r_{v}} \right )} - \nonumber \\  
- \arccos{\frac{\tilde{E}_{n+}}{|\Delta|}}=\pi n,
\end{eqnarray}
where $n\in\{0,\pm 1, \pm 2,\dots$\}, $r_{v}$ denotes radius of the vortex core, and $|L_{z}|=\rho\kF$.
Note that only the states with $n=0$ correspond to core states, i.e.~$E_{\pm, n=0}\lesssim |\Delta|$.
The limit $\frac{|E|}{|\Delta|}\ll 1$ can be quite accurately approximated by the expression:
\begin{equation} \label{andreev_pol}
E_{\pm, n=0,m} \approx -\frac{|\Delta|^{2}}{\eF\frac{r_{v}}{\xi}\left (\frac{r_{v}}{\xi} + 1 \right )}  m \mp\frac{\Delta\mu}{2},
\end{equation} 
where $m$ is the magnetic quantum number associated with $L_{z}=\hbar m$, pointing along the vortex axis and $\xi=\frac{\eF}{\kF |\Delta|}$ is a coherence length.
The energy of the first Andreev state in spin-symmetric case ($\Delta\mu=0$), known as the minigap, is recovered when taking $r_v=\xi$: $E_{0}=|\Delta|^{2}/2\eF$~\cite{volovik2008}.
Since the vortex rotates counterclockwise (generating the flow with positive angular momentum along z-axis $L_{z}>0$) for an unpolarized vortex ($\Delta\mu=0$), the negative energies $E_{\pm,n=0,m}$ correspond
to quasiparticles rotating in the same direction.
In the case of nonzero spin imbalance, the two degenerate branches, corresponding to different spins, become shifted with respect to each other by the value $\Delta\mu$.
Consequently, part of the branch of majority spin particles corresponds to states with the opposite value of $L_{z}$.
The condition $E\approx 0$ sets the limit for the maximum value of the opposite angular momentum generated by the majority spin particle:
\begin{equation}
\max|m_{\textrm{opposite}}|\approx \frac{1}{2}\frac{\eF}{|\Delta|^2}\frac{r_{v}}{\xi}\left (\frac{r_{v}}{\xi} + 1 \right )\Delta\mu.
\end{equation}
In Fig.~\ref{fig:andreev-states}(a) we present comparison of Eq.~(\ref{andreev_pol}) originated from Andreev approximation
and results of direct numerical solution of BdG equations
(see sec.~\ref{sec:suppeq}) for the BCS regime. 
Clearly, the formula reveals satisfactory agreement with data when parameter $r_v$ is set to be approximately $\xi$.

The reversed total flow arises due to the cancellation effect of negative and positive contributions to angular momentum $L_{z}$ inside the core.
To demonstrate this let us consider first the spin symmetric system.
In panel b) of Fig.~\ref{fig:andreev-states} we present contribution to the current in the core ($r\lesssim\xi$) coming from states forming chiral branch only ($E_{\pm, n=0,m}<\Delta$) and contrast it with the total current.
For a better visibility, the currents are expressed in units of maximal total current.
One may notice that contribution from the chiral band already exceeds the total current, which is due to the fact that they are formed by the states having angular momenta coinciding with the vortex, and therefore are the closest to the Fermi surface.
Whereas the states with other values of angular momenta are shifted up in energy.
Thus the current arising from the higher energy states, $E_n>\Delta$, must have reverted circulation.

In the case of spin-imbalanced scenario occupation of the states with opposite angular momentum in the core practically cancels current arising from their positive counterparts~\footnote{To be precise the cancellation is not exact and a small reversed current is produced due to larger occupations of states with opposite angular momentum.This is a consequence of the opposite slope of the bands corresponding to positive and negative angular momenta.}. 
Since states with small $m$ are localized close to the core, the net current carried by the chiral states almost vanishes there, see panel c) of Fig.~\ref{fig:andreev-states}.
In this way, the reversed current produced by non-Andreev states is revealed.
To some extent, the cancellation effect observed here is similar to an effect resulting with the reversion of a supercurrent in a controllable Josephson junctions~\cite{Baselmans,WendinShumeiko,Chang97}, which is due to the occupation pattern of Andreev states.

We note also that, qualitatively, the same effect of reversed circulations is observed in a strongly interacting regime, with the only difference that the density of Andreev states is lower in this case (see discussion of Fig.~\ref{fig:flat-bands}).
The calculations for strong interactions (unitary regime) were carried out within ASLDA framework~\cite{LNP__2012}.
The ASLDA calculations were also conformed with experimental data~\cite{ZwierleinVortexImbalanced} revealing remarkable agreement, and indicating that the vortices with polarized cores were already created in the laboratory~\cite{Kopycinski}.
We point out that the effect gets stronger as we tune the interaction strength towards the deep BCS regime. 
For example for $a\kF\approx-0.6$ the reversed flow in the core has the magnitude comparable to the maximum value of the current outside the core, see Fig.~\ref{fig:supplement}(d).
One has to emphasize that increasing spin polarization even more may eventually lead to spatial modulation of  the order parameter, even in the core, which represent a qualitatively different regime~\cite{Inotani}.

\section{Flat bands and effective mass}
The straight vortex admits the solution in the form of plane waves along the vortex line, which we choose to be the $z$-axis: $\varphi(x,y)e^{ik_z z}$.
A peculiarity of Andreev reflection, however, leads to the significant suppression of the motion along the vortex core.
In the pure Andreev scheme, the quasiparticle at the Fermi surface is reflected exactly backward and thus, except the case of a particle moving exactly in the direction of the vortex line, it will be localized not only within a plane perpendicular to the vortex core, but also along the vortex line.
The manifestation of this behavior will result in almost flat bands $E(k_{z})\propto \textrm{const}$ for $k_{z}\ll\kF$.
This oversimplified result is however modified by the fact that the particles within a core are not exactly at the Fermi surface and, in particular, the departure from the Fermi energy by the value of the minigap $E_0$ leads to a creeping motion along the vortex line.
In this case the particle moving within the core are subject to the Andreev reflection law: $\sqrt{\eF+E}\sin\alpha = \sqrt{\eF-E}\sin\beta$, where $\alpha$ and $\beta$ are angles of trajectories for incident particle and reflected hole respectively \cite{andreev1964,adagideli2002} and $E$ is the quasiparticle energy.
Considering a series of Andreev reflections in a tube of radius $r_v=\xi$, we derive the relation between effective velocity of a particle/hole and momentum $k_z$ along the vortex line which reads (see Appendix~\ref{appendixA} for details):
\begin{equation}
v_{z}=k_{z}\frac{\sqrt{k_{p}^{2}-k_{z}^2}-\sqrt{k_{h}^{2}-k_{z}^2}}{\sqrt{k_{p}^{2}-k_{z}^2}+\sqrt{k_{h}^{2}-k_{z}^2}},
\end{equation}
where $k_{p}=\sqrt{2(\eF + E)}$ and $k_{h}=\sqrt{2(\eF - E)}$.
The above formula estimates the relation for the effective mass of the particle along the vortex axis.
Namely, considering the linear term in $k_{z}$ and $E$ on the rhs one gets: $\meff^{-1}\approx E/2\eF$.
Note that this result agrees with the effective mass derived as $\meff^{-1}=\frac{1}{k_{z}}\frac{dE(k_{z})}{dk_{z}}|_{k_{z}=0}$ from the formula for the dispersion relation in the BCS limit: $E(k_{z})=E(0)/\sqrt{1-k_{z}^{2}/(2 \eF)}$~\cite{caroli1964}.
Consequently one may easily estimate the magnitude of the effective mass component along the vortex line corresponding to angular momenta $L_{z}=\hbar m$: $\meff^{-1}(m)\approx\frac{2|m|}{3}\left (\frac{\Delta}{\eF} \right )^{2}$.
In the deep BCS limit, the inverse of the effective mass will be exponentially small since $\Delta/\eF\propto e^{\pi/2a\kF}$, and clearly the departure from the flat band behavior will be significant at the unitarity where $\Delta/\eF\approx 0.5$.

\begin{figure}[t]
   \begin{center}
   \includegraphics[width=\columnwidth]{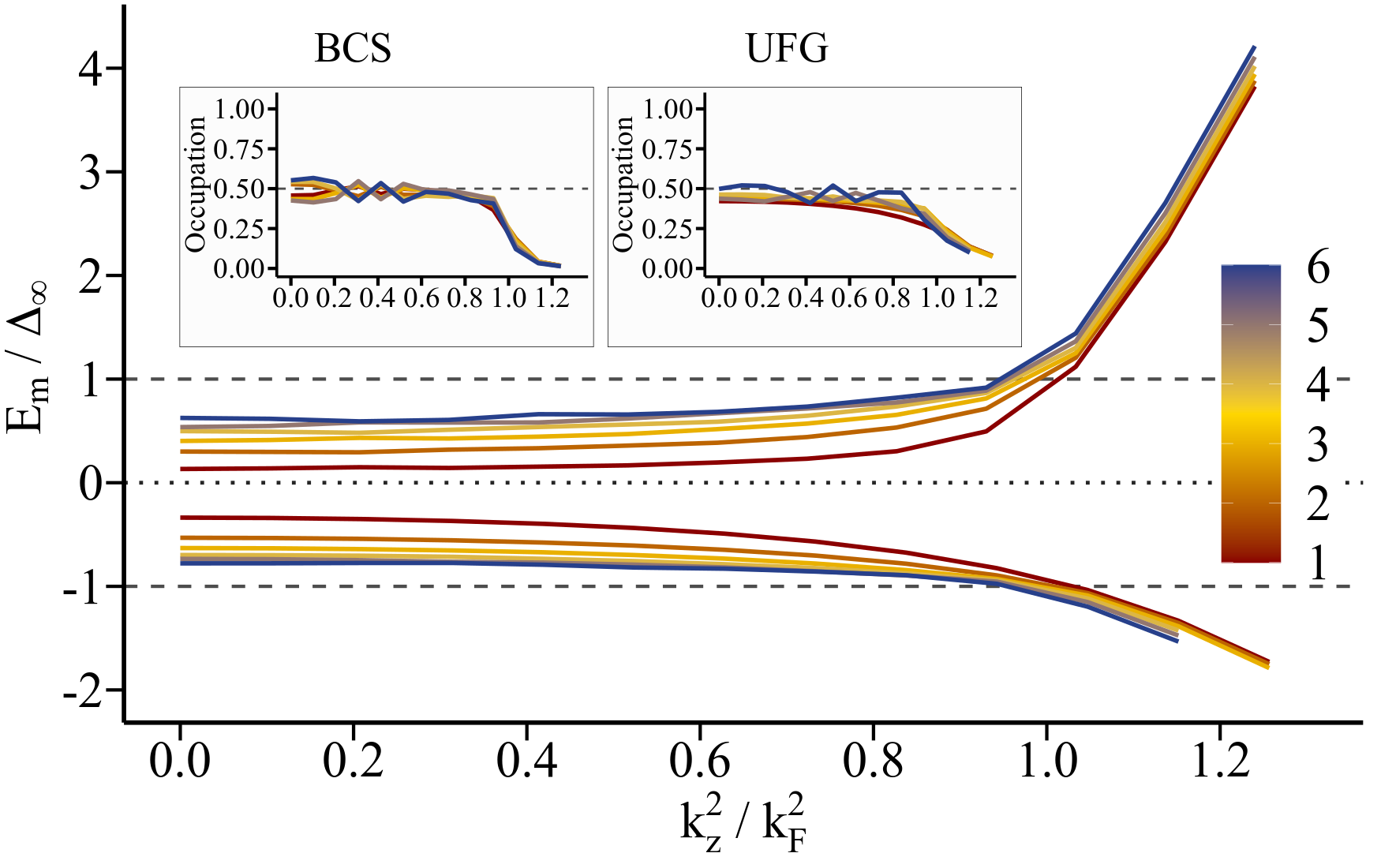} 
   \end{center}\vspace{-3mm}
   \caption{
   Quasiparticle energies of the spin-symmetric system for different $m$-states as a function of momentum $k_z$ along the vortex line.
   In the upper half (positive energies) we show results for the BCS regime ($a\kF=-0.84$), while in the lower half (negative energies) we provide results for the UFG.
   In the insets we display probability of the state to be occupied by the particle $\int |v_{\uparrow/\downarrow}(\textbf{r})|^2 d\bm{r}$, respectively for each case.
   Its value at level $\approx 0.5$ for $k<\kF$ indicates that these states are superpositions of particles and reflected holes, as expected for the Andreev states.
   }
   \label{fig:flat-bands}
\end{figure}

The shape of the bands in the BCS limit and at the unitary regime, are shown in the Fig.~\ref{fig:flat-bands}.
As expected in the BCS regime, the obtained quasienergies form the flat bands for $k<\kF$.
At unitarity, the flatness is less pronounced, which reflects the importance of corrections beyond the Andreev approximation.
In order to demonstrate this fact, let us consider BdG equations for the straight vortex: $H(k_z)\varphi_n(\bm{r}) = E_n(k_z)\varphi_n(\bm{r})$, where $\varphi_n=(u_{n,\uparrow},v_{n,\downarrow})^{T}$ and $\bm{r}=(x,y)$. The Hamiltonian is given as:
\begin{equation} \label{bdg_kz}
 H = \left( \begin{array}{cc}
 h_{2D}(\textbf{r}) + \frac{1}{2}k_{z}^2 -\mu_{\uparrow} & \Delta(\textbf{r}) \\
 \Delta^*(\textbf{r}) & -h_{2D}^{*}(\textbf{r}) - \frac{1}{2}k_{z}^2 + \mu_{\downarrow}
\end{array} \right),
\end{equation}
where $h_{2D}$ describes the 2D part of the single-particle Hamiltonian.
The quasiparticle energy can be computed as $E_n(k_z)=\int \varphi_n^{\dagger}(\bm{r})H\varphi_n(\bm{r})d^{2}\bm{r}$.
Thus, the departure from the flat band behavior is due to the fact that $\frac{dE_{n}}{dk_{z}^{2}}\approx\int \left ( |u_{n}(\textbf{r})|^{2} - |v_{n}(\textbf{r})|^2 \right )d^{2}\bm{r} $, where we have disregarded the dependence of $\Delta$ on $k_{z}$, which is marginal.
Clearly, in the pure Andreev scheme, the integral is exactly zero as the Andreev states are composed of particles and holes in equal proportions.
It is also obvious that the flatness of the band is effective until $\kF$ is reached beyond which $E\propto k_{z}^2$ limit is reproduced.
In the insets of Fig.~\ref{fig:flat-bands}, the occupation probabilities are shown.
The correlation between the departure from the occupation number $1/2$ and the shape of the band is clearly visible.

The band flatness resulting in the increase of the effective mass is going to affect the propagation of the confined polarization along the vortex core.
Namely, in the case of inducing locally polarization of the core, which may occur e.g.~during the reconnection or collision with a polarized vortex~\cite{tylutki2021universal}, it will propagate along the vortex line.
If the local polarization is essentially of $1$-quasiparticle nature, the excitations of the pairing field and spin-waves can be neglected.
Consequently, the propagation along the vortex line will occur simply due to the motion of the wave packet composed of Andreev states carrying spin excess particles.
The propagation will thus occur with velocity $v_{z}=k_{0}/\meff\propto k_{0}\left (\frac{\Delta}{\eF} \right )^{2}$, where $k_{0}$ is the initial momentum of the wave packet.
Similarly the wave packet width in the limit of long times behaves as $\sqrt{\langle (z - v_{z}t)^{2} \rangle} \propto t\left (\frac{\Delta}{\eF} \right )^{2}$ and leads to an effective suppression of the polarization propagation, see Appendix~\ref{AppendixB} for the full derivation.

\section{Classification of polarized vortices}
\begin{figure}[t]
   \begin{center}
   \includegraphics[width=\columnwidth]{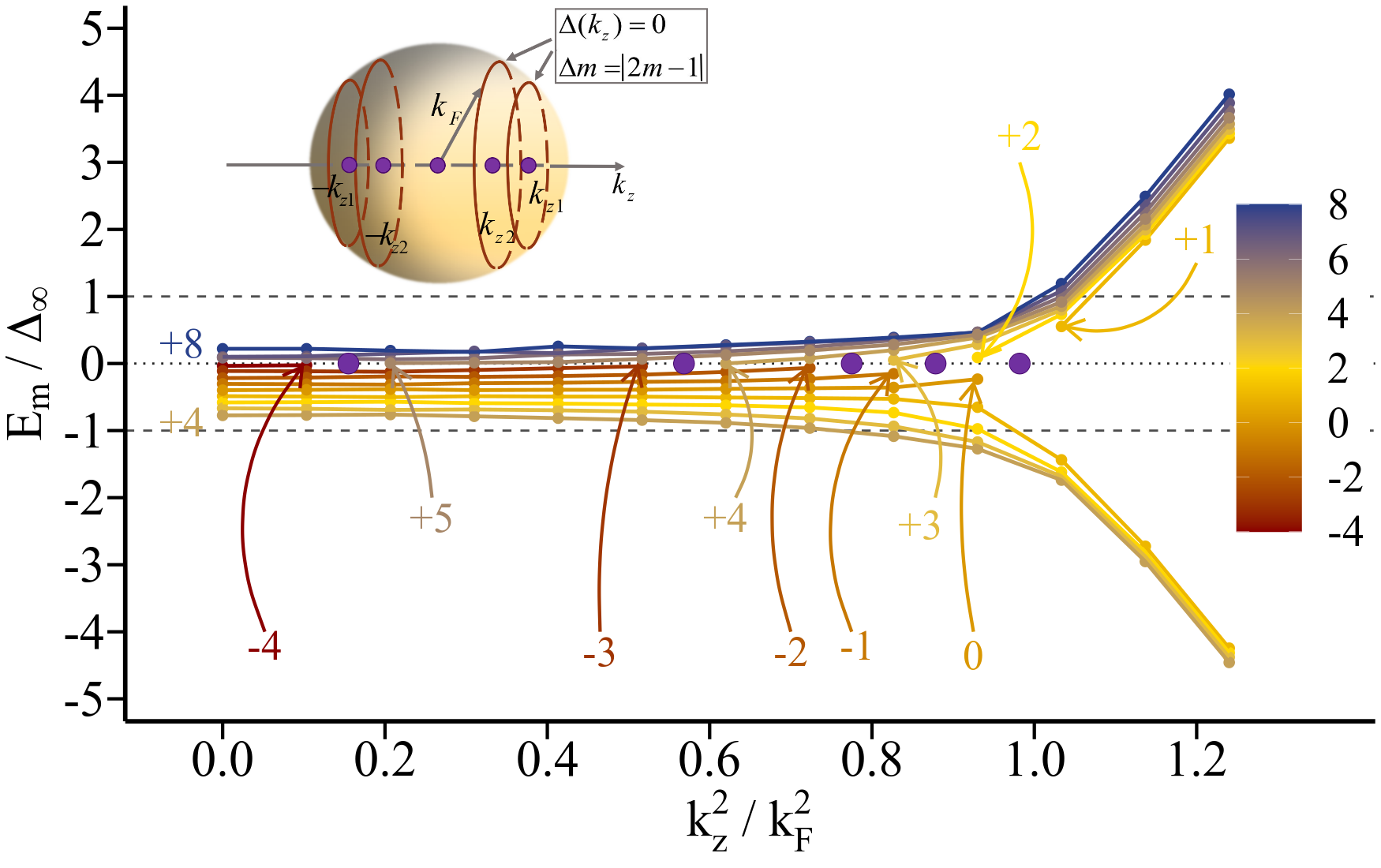} 
   \end{center}\vspace{-3mm}
   \caption{
   Quasiparticle energies of states corresponding to different $m$-values as a function of momentum along the vortex line $k_z$.
   The results are obtained for spin-imbalanced system with $\Delta\mu=0.14\,\eF$ in the BCS regime $a\kF=-0.84$.
   The purple dots on $E = 0$ axis indicate positions of a level crossing, where the configuration changes by $\Delta m=|2m-1|$.
   For a better visibility, the negative energy states of $E_{-}$-branch are shown as positive and the energy levels are not plotted around the crossing point.
   The inset presents the classification of the polarized vortices based on the number of Fermi circles where the minigap vanishes.
   }
   \label{fig:vortex-classification}
\end{figure}

The presence of the polarization in the vortex core leads inevitably to disappearance of the minigap at certain points of the Fermi surface.
It can be seen by examining the spectrum of the Hamiltonian~(\ref{bdg_kz}).
The spin-imbalance generates the relative shift of states corresponding to different spins, and thus the spectrum is not symmetric with respect to $E=0$ and has a different number of positive and negative energy states.
On the other hand, in the limit of large momentum component $k_z^{2}\gg\mu_{\uparrow,\downarrow}$, the spectrum becomes fully symmetric with the same number of positive and negative eigenvalues.
Therefore, one may infer that for certain values of $k_{z}=\pm k_{z1}, \pm k_{z2},\dots$ the spectrum will contain the zero eigenvalues $E(\pm k_{zi})=0$, which correspond to the quasiparticle configuration change.
Precisely, when changing the energy from negative to positive, the particle state $v_{\uparrow}$ with momentum $m$ is converted into hole state $u_{\uparrow}$ with momentum $-m+1$, i.e.~the state that rotate in opposite direction and is shifted by unit of angular momentum with respect to $v_{\downarrow}$ state.
Thus, at the crossing the configuration change by $\Delta m = |2m-1|$ occurs.
This effect is presented on Fig.~\ref{fig:vortex-classification} for spin-imbalanced Fermi gas in the BCS regime.

Since the number of quasiparticle crossings is well defined for a polarized vortex, one can use the number of crossings through the $E=0$ level to classify the vortices in spin-polarized Fermi systems.
Namely, for spin-symmetric vortex, the number of crossings is zero.
Polarizing the vortex is equivalent to introducing a series of crossing at the Fermi surface i.e.~points for which minigap vanishes.
As a consequence, the Fermi sphere will acquire a peculiar structure, consisting of rings which separate regions differing by a peculiar quasiparticle excitation pattern, see inset of Fig.~\ref{fig:vortex-classification} for illustration.

\section{Summary}
We have shown that polarized vortices in Fermi superfluid acquire a peculiar structure with a reversed circulation inside the core.
Their structure admits the vanishing minigap with a characteristic pattern of single-quasiparticle level crossings at the Fermi surface.
It is also predicted that the dynamics along the vortex line of spatially localized polarization inside the core will be suppressed.
Bragg spectroscopy technique may provide experimental signatures of reversed flow~\cite{BraggBdG,Blakie2000}, see also Appendix~\ref{AppendixC}.

\begin{acknowledgments}
We are grateful to Michael McNeil Forbes for reading the manuscript and various suggestions.
This work was supported by the Polish National Science Center (NCN) under
Contracts No.UMO-2016/23/B/ST2/01789 (AM), UMO-2017/27/B/ST2/02792 (PM,KK) and UMO-2017/26/E/ST3/00428 (GW).
We acknowledge PRACE for awarding us access to resource Piz Daint
based in Switzerland at Swiss National Supercomputing Centre (CSCS), decision No.2019215113.
We also acknowledge Computational Modelling (ICM) of Warsaw University for computing resources at Okeanos (grant No.GA67-14) and PL-Grid Infrastructure for providing us resources at Prometheus supercomputer.
\end{acknowledgments}

\appendix
\section{Andreev states in the core of polarized vortex}\label{appendixA}

\begin{figure}[h]
   \begin{center}
   \includegraphics[width=0.35\textwidth]{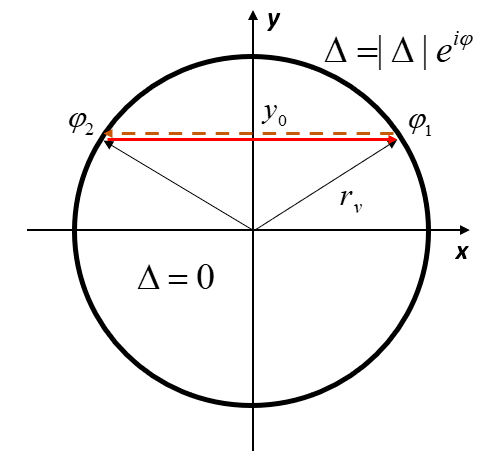}
   \centering
   \end{center}\vspace{-3mm}
   \caption{Schematic picture of the vortex core used for determination of states in Andreev approximation. 
   The classical trajectory representing particle of momentum $k_{F}$ is denoted by red solid line, and
   reflected hole is shown as brown dashed line. Note that the angular momentum component $L_{z}$ corresponding 
   to the trajectory is negative.}
   \label{fig:supplement1}
\end{figure}

The Andreev approximation assumes separation of two length scales: $\kF^{-1}\ll\xi$
($\xi=\frac{\eF}{\kF |\Delta|}$ being coherence length). It clearly 
holds in deep BCS regime and 
then also $\mu\approx \eF=\frac{\hbar^2\kF^2}{2m}$ is satisfied. The components of quasiparticle wave-functions attain generic form $\varphi(\bm{r})=e^{i\bm{k}_\textrm{F}\cdot\bm{r}}\tilde{\varphi}(\bm{r})$. Action of the hamiltonian $(h-\mu)\varphi$ simplifies to:
\begin{equation}
\left(-\frac{\hbar^2}{2m}\nabla^2-\mu\right)\varphi(\bm{r})
\approx
e^{i\bm{k}_\textrm{F}\cdot\bm{r}}\left(
-\frac{i\hbar}{m}\bm{k}_\textrm{F}\cdot\grad\tilde{\varphi}(\bm{r})
\right),
\label{eqn:andreev-hamiltonian}
\end{equation}
where the term proportional to $\nabla^2\tilde{\varphi}$ 
is neglected, due to assumption of slow variation of the function $\tilde{\varphi}$ over the length scale $\kF^{-1}$.  Inserting (\ref{eqn:andreev-hamiltonian}) into (\ref{eq:hfbspin1}) one arrives at Eq.~(\ref{bdg1}) from 
the main paper (we set units: $\hbar=m=1$):
\begin{equation} \label{bdg1-sm}
 \left( \begin{array}{cc}
 -i\bm{k}_\textrm{F}\cdot\grad& \Delta(\textbf{r}) \\
 \Delta^*(\textbf{r}) & i\bm{k}_\textrm{F}\cdot\grad
\end{array} \right)\left( \begin{array}{cc}
\tilde{u}_{n,\uparrow}(\textbf{r}) \\
\tilde{v}_{n,\downarrow}(\textbf{r})
\end{array} \right) =\tilde{E}_{n+} \left( \begin{array}{cc}
\tilde{u}_{n,\uparrow}(\textbf{r}) \\
\tilde{v}_{n,\downarrow}(\textbf{r})
\end{array} \right),
\end{equation}
where $\tilde{E}_{n+} = E_{n+} + \frac{\Delta\mu}{2}$.
The second pair of equations for $\tilde{u}_{n,\downarrow}(\textbf{r})$ and $\tilde{v}_{n,\uparrow}(\textbf{r})$ has similar form and correspond to $\tilde{E}_{n-} = E_{n-} - \frac{\Delta\mu}{2}$:
\begin{equation} \label{bdg2}
 \left( \begin{array}{cc}
 -i\bm{k}_\textrm{F}\cdot\grad& -\Delta(\textbf{r}) \\
 -\Delta^*(\textbf{r}) & i\bm{k}_\textrm{F}\cdot\grad
\end{array} \right)\left( \begin{array}{cc}
\tilde{u}_{n,\downarrow}(\textbf{r}) \\
\tilde{v}_{n,\uparrow}(\textbf{r})
\end{array} \right) =\tilde{E}_{n-} \left( \begin{array}{cc}
\tilde{u}_{n,\downarrow}(\textbf{r}) \\
\tilde{v}_{n,\uparrow}(\textbf{r})
\end{array} \right),
\end{equation}
In the case of the schematic vortex core structure in the form
$\Delta(\bm{r}) = \Delta(\rho,\phi)=|\Delta|e^{i\phi}\theta(\rho-r_{v})$ (counterclockwise rotating vortex) one arrives at the quantization condition from eq.~(\ref{bdg1-sm}) (see Fig.~
\ref{fig:supplement1}):
\begin{eqnarray} \label{quant}
\frac{\tilde{E}_{n+}}{\eF}\kF r_{v}\sqrt{1-\left (\frac{y_{0}}{r_{v}} \right )^{2}}+\arccos{\left ( \frac{y_{0}}{r_{v}} \right )} - \nonumber \\  
- \arccos{\frac{\tilde{E}_{n+}}{|\Delta|}}=\pi n,
\end{eqnarray}
where $n=0,\pm 1, \pm 2, ...$. Introducing the angular momentum component $L_{z}=-y_{0}\kF$
one gets:
\begin{eqnarray} \label{quant1}
\frac{\tilde{E}_{n+}}{\eF}\kF r_{v}\sqrt{1-\left (\frac{L_{z}}{\kF r_{v}} \right )^{2}}+\arccos{\left ( \frac{-L_{z}}{\kF r_{v}} \right )} - \nonumber \\  
- \arccos{\frac{\tilde{E}_{n+}}{|\Delta|}}=\pi n.
\end{eqnarray}
In the limit of $|y_{0}/r_{v}|\ll 1$ and $|\tilde{E}_{n+}/\Delta |\ll 1$ the equation
simplifies to:
\begin{eqnarray} \label{eplus}
\tilde{E}_{+} \approx -\frac{|\Delta|^{2}}{\eF\frac{r_{v}}{\xi}\left (\frac{r_{v}}{\xi} + 1 \right )}  L_{z},
\end{eqnarray}
where only the lowest energy branch corresponding to $n=0$ is considered. 
Note that the minus sign appears as a result of counterclockwise rotation of the superflow. 
The other solution corresponding to eq.~(\ref{bdg2}) can be obtained 
by noting
that they are equivalent to complex conjugate solutions of (\ref{bdg1-sm}) with relation $E_{-}=-E_{+}$. Therefore the particle momentum is reversed 
$\bm{k}_\textrm{F}\rightarrow -\bm{k}_\textrm{F}$ and consequently 
$L_{z}\rightarrow -L_{z}$. As a result one arrives at $\tilde{E}_{-}=\tilde{E}_{+}$. 
In the case of spin-polarized system the solutions are shifted with respect to each other
by $\Delta\mu$. 
Note that the energies corresponding to the highest angular momenta in the core are of the 
order of $|\Delta|$.
Namely, for the maximum $L_{z}=\mp\kF r_{v}$ one gets 
$\tilde{E}_{\pm}=\pm |\Delta|$, respectively. 

\begin{figure}[h]
   \begin{center}
   \includegraphics[width=0.4\textwidth]{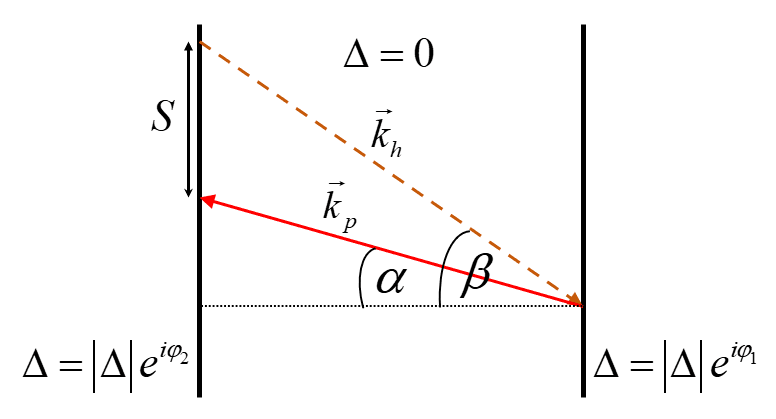}
   \centering
   \end{center}\vspace{-3mm}
   \caption{Schematic picture of the section through the vortex core used for 
    determination of the effective mass along the vortex line. 
   The classical trajectory representing particle of momentum $k_{p}$ is denoted by red solid line, and
   reflected hole of momentum $k_{h}$ is shown as brown dashed line.}
   \label{fig:supplement2}
\end{figure}

In order to extract the effective mass in the Andreev approximation one needs to consider 
particle/hole motion along the vortex line. Due to the properties of Andreev reflection the problem reduces
to 2D problem, see Fig.~\ref{fig:supplement2}. Contrary to the quantization condition which
resulted from the assumption that the hole(particle) is reflected exactly backward (which
is true if the incoming particle(hole) is exactly at the Fermi surface), here
one needs to take into account more general case. 
Namely, as a result of momentum conservation along the vortex line the reflection law reads: 
$\sqrt{\eF + E}\sin\alpha = \sqrt{\eF - E}\sin\beta$, where $k_{p}=\sqrt{2(\eF + E)}$ and
$k_{h}=\sqrt{2(\eF - E)}$, are particle and hole momenta, respectively. The effective velocity
along the vortex line can be defined as $v=S/T$, where $S$ denotes the distance between two consecutive
reflections where particle is converted into hole (see Fig.~\ref{fig:supplement2}), and
$T$ is the time interval between these reflections. Consequently one gets:
$v=\sqrt{2(\eF + E)}\sin\alpha\sin(\beta-\alpha)/\sin(\beta+\alpha)$. 
Using the reflection law this relation can be rewritten as:
\begin{equation}
v_{z}=k_{z}\frac{\sqrt{k_{p}^{2}-k_{z}^2}-\sqrt{k_{h}^{2}-k_{z}^2}}{\sqrt{k_{p}^{2}-k_{z}^2}+\sqrt{k_{h}^{2}-k_{z}^2}},
\end{equation}
where $k_{z}=k_{p}\sin\alpha=k_{h}\sin\beta$ is the momentum component along the vortex line. 
Note that the expression does not depend on the core radius and therefore in the Andreev approximation
all bands originated from states (\ref{eplus}) will have the same slope. 
Andreev approximation in practice is expected to work for small $|L_{z}|\ll \kF r_v$ and small 
$k_{z}\ll \kF$ (small angles of reflection) as is shown in the manuscript. 

\section{Wave packet excitation in the vortex core}\label{AppendixB}

Let us consider an unpolarized vortex of length $L$. 
The Hamiltonian describing the structure of the vortex
core reads:
\begin{eqnarray}
\hat{H} &=& \frac{L}{2\pi}\int dk_{z}\sum_{m>0} \Bigg [  
E_{m\uparrow}(k_{z})\alpha_{m\uparrow}^{\dagger}(k_{z}) \alpha_{m\uparrow}(k_{z}) + \nonumber \\
& &E_{m\downarrow}(k_{z})\alpha_{m\downarrow}^{\dagger}(k_{z}) \alpha_{m\downarrow}(k_{z}) 
\Bigg ]
\end{eqnarray}
where for $k_{z}/k_F \ll 1$: $E_{m\uparrow\downarrow}(k_{z}) \approx \Omega m + \frac{1}{2\meff}k_{z}^2$ with $\Omega$ being proportionality coefficient between energy and quantum number $m=L_z/\hbar$ in Eq.~(\ref{eplus}) and
\begin{eqnarray}
& &\alpha_{m\uparrow\downarrow}^{\dagger}(k_{z})=  \\
& &\int d^{3}r e^{i k_{z} z}\left ( v_{m}(\rho)e^{i m \phi}a_{\downarrow\uparrow}({\bf r}) +
                    u_{m}(\rho)e^{i (m-1) \phi}a_{\uparrow\downarrow}^{\dagger}({\bf r})
                    \right ) \nonumber .
\end{eqnarray}
One quasiparticle excitation within a band formed by states with well defined $m$-value
can be constructed in the standard way:
\begin{eqnarray}
& &|k_{0} m \uparrow\downarrow\rangle = \\
& &
\frac{1}{\sqrt{\sqrt{2\pi}\sigma}}\int dk_{z} \exp\left (-\frac{(k_{z}-k_{0})^{2}}{4\sigma^{2}} \right )
\alpha_{m\uparrow\downarrow}^{\dagger}(k_{z})|0\rangle  \nonumber
\end{eqnarray}
and clearly $\langle k_{0} m \uparrow\downarrow|k_{0} m \uparrow\downarrow\rangle =1$. 
The wave packet excitation change the spin polarization by unity, since eg. 
$\langle k_{0} m \uparrow|(\hat{N}_{\uparrow}-\hat{N}_{\downarrow} )| k_{0} m \uparrow\rangle
-\langle 0|(\hat{N}_{\uparrow}-\hat{N}_{\downarrow} )|0\rangle = 1$, where 
$\hat{N}_{\uparrow}$, $\hat{N}_{\downarrow}$ are particle number operators for spin-up and spin-down
partcile, respectively. 
The evolution of this wave packet: 
$|k_{0} m \uparrow\downarrow, t\rangle = \exp(-i \hat{H} t) |k_{0} m \uparrow\downarrow\rangle$
gives rise to the relations:
\begin{equation}
\langle z \rangle = \langle k_{0} m \uparrow\downarrow|z|k_{0} m \uparrow\downarrow\rangle = 
\frac{k_{0}}{\meff} t
\end{equation}
\begin{equation}
\sqrt{\langle \left (z - \frac{k_{0}}{\meff} t \right )^{2}\rangle } = 
\frac{1}{2\sigma}\sqrt{\left (2 \frac{\sigma^{2}}{\meff} t \right )^2+1}
\approx \frac{\sigma}{\meff} t
\end{equation}
for long times: $t\gg \frac{\meff}{2\sigma^{2}}$.

\section{Impact of reversed circulation on Bragg scattering}\label{AppendixC}

\begin{figure*}[]
~\textbf{a)} BCS 0\%, $a k_F$ = -0.61
~~~~~~~~~~~~~~~~~~~~~~~~~~~~~~~~~~~~~~~~~~~~~~~
 \textbf{b)} BCS 0.5\%, $a k_F$ = -0.61
~~~~~~~~~~~~~~~~~~~~~~~~~~~~~~~~~~~~~~~~~~~~\\
   \begin{center}
   \includegraphics[width=0.40\textwidth]{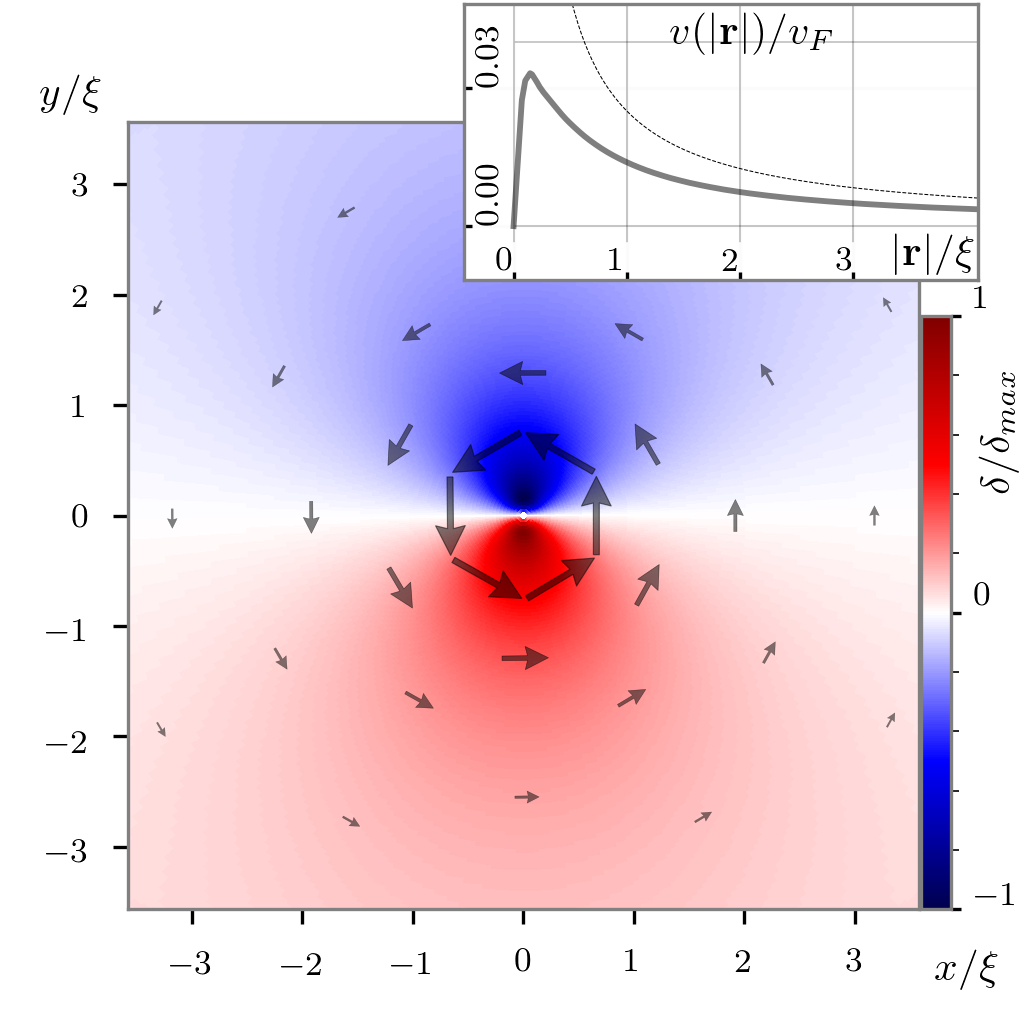}
   \includegraphics[width=0.40\textwidth]{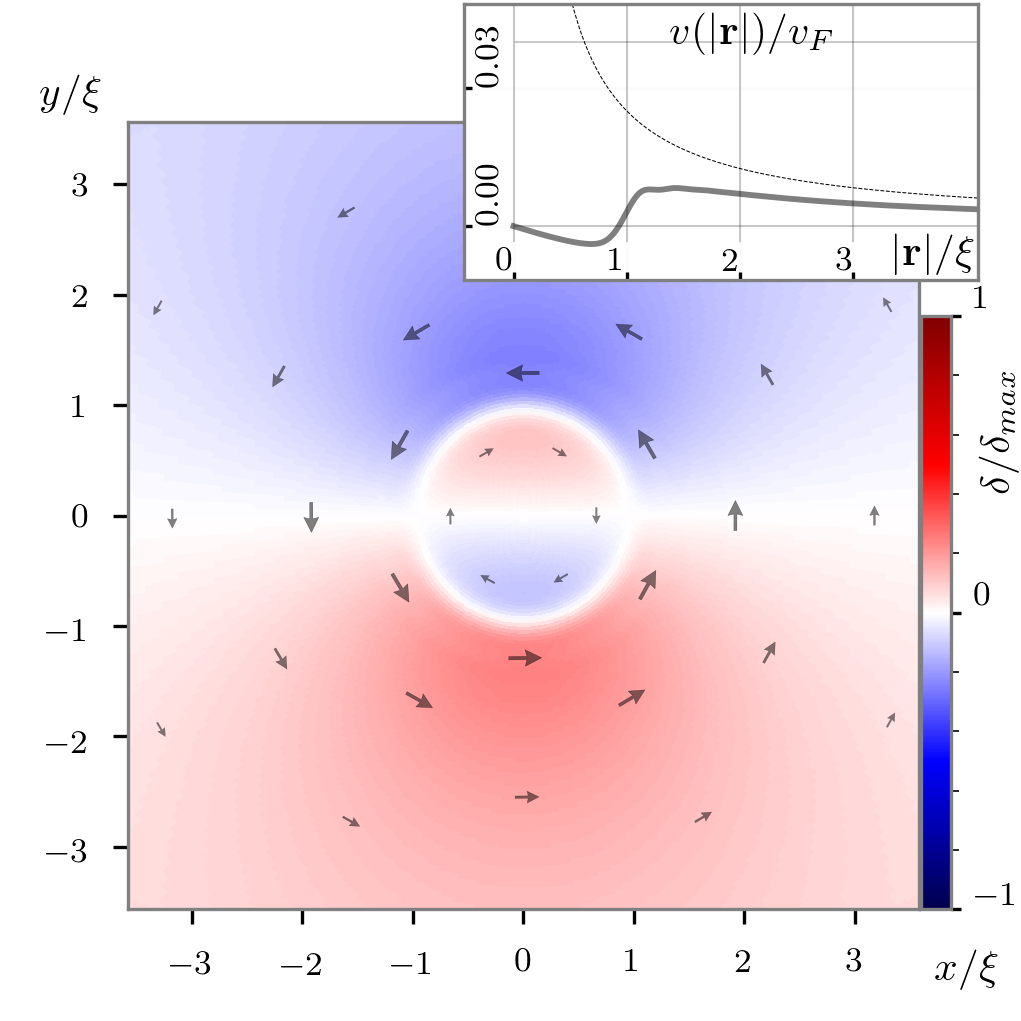} 
   \end{center}\vspace{-3mm}
\caption{Color maps are showing the relative change in Bragg scattering resonance frequency distribution $\delta(\bm{r})/ \delta_{max}$ due to velocity of atoms for spin unpolarized (BCS, P=0\%) and spin polarized systems (BCS, P=0.5\%), in panels (a) and (b), respectively. In this particular setup
we chose $\bm{q}$ to be aligned along $x$-axis of the system (while vortex is oriented along $z$-axis). 
The quantity is normalized to its maximal value $\delta_{max}$ for unpolarized case. Vector field related to the vortex $v(\bm{r})$ is indicated by arrows. In the insets we show corresponding velocity profiles as a function of distance from the vortex core obtained numerically (solid line). The ideal quantum vortex velocity profile $v(r)\sim 1/r$ is marked by dashed line.}
   \label{fig:supplement_Bragg}
\end{figure*}

Reversed circulation is manifested as a change in the collective motion of atoms in a condensate. Bragg spectroscopy can be a promising tool for the investigation of this effect. Below we present qualitative arguments supporting the design of the Bragg scattering experiment, omitting the issue if current experimental capabilities allow sufficiently accurate measurements.

Bragg scattering experiments were successfully employed to investigate fermionic condensates~\cite{Veeravalli2008,Lingham2016} as well as to probe quantum vortices in BEC~\cite{Muniz2006,Seo2017}. In a typical setup of the experiment two laser beams (having certain frequency difference $\omega$) are generated,
crossing each other inside the atomic cloud. 
They produce a standing wave moving in the laboratory frame and 
thus inducing Bragg scattering of the atomic cloud. 
Namely, crossing laser beams form an effective optical potential $V_{opt} \propto \cos\left(\bm{q}\cdot\bm{r} - \omega t\right)$ acting on a gas~\cite{PitaevskiiStringari2001,BraggBdG}. As a result, energy $\hbar\omega$ and momentum $\bm{q}$ are transferred to an atom through the two-photon scattering process.

The resonant Bragg scattering occurs under condition:
\begin{equation} \label{eq:BraggResonanceCondition}
\hbar \omega = \frac{\hbar^2 \bm{q}^2}{2m} + \frac{\bm{q}\cdot\bm{v}}{\hbar},
\end{equation}
where $\bm{v}$ denotes velocity of an atom. 
In the above expression we assumed that the dispersion relation
for an atom in the cloud is the same as for non-interacting particle
(see e.g.~\cite{Lingham2016,Muniz2006,Blakie2000}), although
more realistic expression can be employed as well. 
The second term is crucial in this case as it makes Bragg
scattering process sensitive to local atomic velocity.
In the case of ultracold Fermi gas with vortex line, we define the velocity field through ratio of the probability current and the density $\bm{v}(\bm{r}) = \bm{j}(\bm{r}) / n(\bm{r})$, which corresponds to expectation value of single atom velocity.
Note that Bragg scattering process selects in this case group of atoms from a particular part of the system where the condition holds:
\begin{equation} \label{eq:BraggResonanceCondition1}
\hbar (\omega-\delta(\bm{r})) = \frac{\hbar^2 \bm{q}^2}{2m}.
\end{equation}
with $\hbar\delta(\bm{r}) = \bm{q}\cdot\bm{v}(\bm{r}) / \hbar$.
The quantity $\delta(\bm{r})$ is shown in Fig.~\ref{fig:supplement_Bragg} for vortex with and without reversed flow. 
The figure reveals qualitative and quantitative changes of resonant frequency distribution due to the reversed circulation.

As an experimental signal one can use density distribution of scattered atoms \cite{Veeravalli2008,Muniz2006,Seo2017}.
Due to the sensitivity of Bragg scattering process on the
local flow velocity the presence of reversed circulation should 
induce a significant modification in the density distribution.
Consequently we expect that the density distributions corresponding to
spin unpolarized and polarized vortex can be distinguished.
We emphasize that more refined study of Bragg scattering intensity and
the density distribution evolution for given $\mathbf{q},\omega$ is required in order to settle if such measurements are feasible.

\bibliographystyle{apsrev4-1}
\input{bibl.bbl}

\end{document}

%% file: bibl.bbl
%